\documentclass[a4paper,12pt]{article}
\usepackage{amssymb}

\begin{document}

\title{Varying Cosmological Constant and the Machian Solution
 in the Generalized Scalar-Tensor Theory}
\author{A. Miyazaki \thanks{
Email: miyazaki@loyno.edu, miyazaki@nagasakipu.ac.jp} \vspace{3mm} \\
\textit{Department of Physics, Loyola University, New Orleans, LA 70118} \\
and \\
\textit{Faculty of Economics, Nagasaki Prefectural University} \\
\textit{Sasebo, Nagasaki 858-8580, Japan}}
\date{\vfill}
\maketitle

\begin{abstract}
The cosmological constant $(1/2)\lambda _{1}\phi _{,\,\mu }\phi ^{,\,\mu
}/\phi ^{2}$ is introduced to the generalized scalar-tensor theory of
gravitation with the coupling function $\omega (\phi )=\eta /(\xi -2)$ and
the Machian cosmological solution satisfying $\phi =O(\rho /\omega )$ is
discussed for the homogeneous and isotropic universe with a perfect fluid
(with negative pressure). We require the closed model and the negative
coupling function for the attractive gravitational force. The constraint $%
\omega (\phi )<-3/2$ for $0\leqq \xi <2$ leads to $\eta >3$. If $\lambda
_{1}<0$ and $0\leqq -\eta /\lambda _{1}<2$, the universe shows the slowly
accelerating expansion. The coupling function diverges to $-\infty $ and the
scalar field $\phi $ converges to $G_{\infty }^{-1}$ when $\xi \rightarrow 2$
($t\rightarrow +\infty $). The cosmological constant decays in proportion to 
$t^{-2}$. Thus the Machian cosmological model approaches to the Friedmann
universe in general relativity with $\ddot{a}=0$, $\lambda =0$, and $p=-\rho
/3$ as $t\rightarrow +\infty $. General relativity is locally valid enough
at present. \newline
\end{abstract}

\newpage

Since the generalized scalar-tensor theory of gravitation \cite{1)}-\cite{3)}
was proposed, many attempts to determine the arbitrary coupling function $%
\omega (\phi )$ have been made (see, for example, \cite{4)}-\cite{9)} and
references therein). Recently we discussed a new aspect of the coupling
function in the Machian point of view and proposed $\omega (\phi )=\eta
/(\xi -2)$ \cite{10)}. This coupling function does not include explicitly
the scalar field $\phi $ but depends on the parameter $\xi $, and varies in
time very slowly owing to the physical revolution of matter in the universe (%
$0\leqq \xi (t)<2$). As the parameter $\xi $ approaches to $2$ (negative
pressure $\gamma =-1/3$), the coupling function diverges to the minus
infinity. If we assume $\epsilon \equiv 2-\xi \sim 10^{-3}$ at present, we
obtain $\omega \sim -10^{3}$ which is compatible with the recent
observations \cite{11)}.

For this coupling function, the generalized scalar-tensor theory of
gravitation has a simple cosmological solution satisfying $\phi =O(\rho
/\omega )$ (Machian solution) for the homogeneous and isotropic universe
with a perfect fluid with pressure. The scalar field in this closed model
converges to a finite constant $G_{\infty }^{-1}>0$ (, which does not differ
much from the present gravitational constant $G_{0}^{-1}$) when $\xi
\rightarrow 2$ (probably, $t\rightarrow +\infty $). So the Machian
cosmological solution realize dynamically the almost constant gravitational
''constant'' as the result of the evolution of the universe.

However, this cosmological model shows the extremely slowly decelerating
expansion when $\xi \rightarrow 2$, which is not compatible with the recent
measurements \cite{12)} of the distances to type Ia supernovae. In the
present paper, we discuss the Machian cosmological solution in the
(modified) generalized scalar-tensor theory of gravitation with the varying
cosmological constant to realize the slowly accelerating expansion of the
universe.

Let us start with the following action \cite{1)}-\cite{3)} 
\begin{equation}
S=\int d^{4}x\sqrt{-g}\left\{ -\phi \left[ R+\lambda (\phi ,\phi _{,\,\mu
}\phi ^{,\,\mu })\right] +16\pi L_{m}-\frac{\omega (\phi )}{\phi }g^{\mu \nu
}\phi _{,\,\mu }\phi _{,\,\nu }\right\} \,,  \label{e1}
\end{equation}
where $R$ is the scalar curvature of the metric $g_{\mu \nu }$, $\phi (x)$
is the Brans-Dicke scalar field, $\omega (\phi )$ is an arbitrary coupling
function, and $L_{m}$ represents the Lagrangian for the matter fields. The
term $\lambda (\phi ,\phi _{,\,\mu }\phi ^{,\,\mu })$ represents the
cosmological constant, which is modified from $\lambda (\phi )$, a function
of only $\phi $. It should be noted that the term $\phi _{,\,\mu }\phi
^{,\,\mu }$ is a scalar. The ordinary cosmological constant must be positive
($\lambda >0$) for the cosmological repulsion. We assume a particular form
of the cosmological constant in the Machian point of view (in order to
realize a solution satisfying $\phi =O(\rho /\omega )$): 
\begin{equation}
\lambda (\phi ,\phi _{,\,\mu }\phi ^{,\,\mu })\equiv \frac{1}{2}\lambda
_{1}\phi _{,\,\mu }\phi ^{,\,\mu }/\phi ^{2}\,,  \label{e2}
\end{equation}
where $\lambda _{1}$ is a constant (later, we require $\lambda _{1}<0$).\
For this particular cosmological constant, we obtain from the action Eq.(\ref
{e1}) 
\begin{equation}
S=\int d^{4}x\sqrt{-g}\left\{ -\phi R+16\pi L_{m}-\frac{\left[ \omega (\phi
)+\lambda _{1}\right] }{\phi }g^{\mu \nu }\phi _{,\,\mu }\phi _{,\,\nu
}\right\} \,.  \label{e3}
\end{equation}
We can regard that the present cosmological constant is a constant-part of
the coupling function of $\phi $. The variation of Eq.(\ref{e3}) with
respect to $g_{\mu \nu }$ and $\phi $ leads to the field equations 
\begin{eqnarray}
R_{\mu \nu }-\frac{1}{2}Rg_{\mu \nu } &=&\frac{8\pi }{\phi }T_{\mu \nu }-%
\frac{\left[ \omega (\phi )+\lambda _{1}\right] }{\phi ^{2}}\left( \phi
_{,\,\mu }\phi _{,\,\nu }-\frac{1}{2}g_{\mu \nu }\phi _{,\,\lambda }\phi
^{,\,\lambda }\right)  \nonumber \\
&&-\frac{1}{\phi }(\phi _{,\,\mu ;\,\nu }-g_{\mu \nu }\square \phi )\,,
\label{e4}
\end{eqnarray}
\begin{equation}
\square \phi =-\frac{1}{3+2\left[ \omega (\phi )+\lambda _{1}\right] }\left[
8\pi T+\frac{d\omega (\phi )}{d\phi }\phi _{,\,\lambda }\phi ^{,\,\lambda }%
\right] \,,  \label{e5}
\end{equation}
which satisfy the conservation law of the energy-momentum $T_{\mu \nu }$%
\begin{equation}
T_{;\nu }^{\mu \nu }=0\,.  \label{e6}
\end{equation}

The line element for the Friedmann-Robertson-Walker metric is 
\begin{equation}
ds^{2}=-dt^{2}+a^{2}(t)[d\chi ^{2}+\sigma ^{2}(\chi )(d\theta ^{2}+\sin
^{2}\theta d\varphi ^{2})]\,,  \label{e7}
\end{equation}
where $\sigma (\chi )$ is $\sin \chi $, $\chi $, and $\sinh \chi $ for
closed ($k=+1$), flat ($k=0$), and open ($k=-1$) spaces, respectively. The
energy-momentum tensor for the perfect fluid with pressure $p$ and the mass
density $\rho $\ is given as 
\begin{equation}
T_{\mu \nu }=-pg_{\mu \nu }-(\rho +p)u_{\mu }u_{\nu }\,.  \label{e8}
\end{equation}
The nonvanishing components are $T_{00}=-\rho $, $T_{i\,i}=-pg_{i\,i}$ ($%
i\neq 0$), and its trace is $T=\rho -3p$ for the homogeneous and isotropic
universe.

The energy conservation Eq.(\ref{e6}) leads to the equation of continuity 
\begin{equation}
\dot{\rho}+3\frac{\dot{a}}{a}\left( \rho +p\right) =0\,,  \label{e9}
\end{equation}
which gives, with the barotropic equation of state 
\begin{equation}
p(t)=\gamma \rho (t)\,,\,-1\leqq \gamma \leqq 1/3\,,  \label{e10}
\end{equation}
\begin{equation}
\rho (t)a^{n}(t)=const\,,  \label{e11}
\end{equation}
where $n=3(\gamma +1)$.

The nonvanishing components of the field equation (\ref{e4}) are 
\begin{eqnarray}
2a\ddot{a}+\dot{a}^{2}+k &=&-\frac{\left[ \omega (\phi )+\lambda _{1}\right] 
}{2}a^{2}\left( \frac{\dot{\phi}}{\phi }\right) ^{2}+a\dot{a}\left( \frac{%
\dot{\phi}}{\phi }\right)  \nonumber \\
&&-\frac{8\pi a^{2}p}{\phi }-\frac{8\pi a^{2}\left( \rho -3p\right) }{3+2%
\left[ \omega (\phi )+\lambda _{1}\right] }\frac{1}{\phi }\,,  \label{e12}
\end{eqnarray}
and 
\begin{equation}
\frac{3}{a^{2}}\left( \dot{a}^{2}+k\right) =\frac{\left[ \omega (\phi
)+\lambda _{1}\right] }{2}\left( \frac{\dot{\phi}}{\phi }\right) ^{2}+\frac{%
\ddot{\phi}}{\phi }+\frac{8\pi \rho }{\phi }-\frac{8\pi \left( \rho
-3p\right) }{3+2\left[ \omega (\phi )+\lambda _{1}\right] }\frac{1}{\phi }\,.
\label{e13}
\end{equation}
The field equation (\ref{e5}) gives 
\begin{equation}
\ddot{\phi}+3\frac{\dot{a}}{a}\dot{\phi}=\frac{1}{3+2\left[ \omega (\phi
)+\lambda _{1}\right] }\left[ 8\pi \left( \rho -3p\right) -\frac{d\omega
(\phi )}{d\phi }\dot{\phi}^{2}\right] \,.  \label{e14}
\end{equation}
We adopt as usual Eqs.(\ref{e13}), (\ref{e14}), and (\ref{e11}) as the
independent equations to solve simultaneously.

Let us require as the coupling function 
\begin{equation}
\omega (\phi )\equiv \frac{\eta }{\xi -2}\,,  \label{e15}
\end{equation}
which is necessarily derived from the condition for the reasonable Machian
solution \cite{10)}. The parameter $\xi =1-3\gamma $ varies in time very
slowly as a quasi-static process and so we may regard that the parameter $%
\xi $ is constant when we execute the derivative with respect to $t$. We
introduce another scalar function $\Phi (t)$ by 
\begin{equation}
\phi (t)=\frac{8\pi }{3+2\left[ \omega (\phi )+\lambda _{1}\right] }\Phi (t)
\label{e16}
\end{equation}
for the Machian solution satisfying $\phi =O(\rho /\omega )$. Taking Eq.(\ref
{e16}) and $d\omega /d\phi =0$ into account, we obtain from Eq.(\ref{e14}) 
\begin{equation}
\ddot{\Phi}+3\frac{\dot{a}}{a}\dot{\Phi}=\xi \rho \,,  \label{e17}
\end{equation}
which means that the ratio $\dot{a}/a$ does not include $\omega $, and so we
find for the expansion parameter 
\begin{equation}
a(t)\equiv A(\omega )\alpha (t)\,,  \label{e18}
\end{equation}
where $A$ and $\alpha $\ are arbitrary functions of only $\omega $\ and $t$\
respectively.

After eliminating $\ddot{\phi}$ by Eq.(\ref{e14}), we get from Eq.(\ref{e13}%
) 
\begin{eqnarray}
&&\frac{\omega }{2}\left[ \left( \frac{\dot{\Phi}}{\Phi }\right) ^{2}+\frac{%
4\rho }{\Phi }\right] -\frac{3k}{A^{2}(\omega )\alpha ^{2}}  \nonumber \\
&=&3\left( \frac{\dot{\alpha}}{\alpha }\right) ^{2}+3\left( \frac{\dot{\alpha%
}}{\alpha }\right) \left( \frac{\dot{\Phi}}{\Phi }\right) -\frac{\lambda _{1}%
}{2}\left( \frac{\dot{\phi}}{\phi }\right) ^{2}-\frac{\left( 3+2\lambda
_{1}\right) \rho }{\Phi }\,.  \label{e19}
\end{eqnarray}
For the closed and the open spaces ($k=\pm 1$), if we require that Eq.(\ref
{e19}) is identically satisfied for all arbitrary values of $\omega $, we
find that\ the coefficient $A(\omega )$ must have the following form 
\begin{equation}
\frac{3}{A^{2}(\omega )}=\left| \frac{\omega (\phi )}{2}+B\right| \,,
\label{e20}
\end{equation}
where $B$\ is a constant with no dependence of $\omega $ and furthermore we
obtain 
\begin{equation}
\left( \frac{\dot{\Phi}}{\Phi }\right) ^{2}+\frac{4\rho }{\Phi }\equiv k\,j%
\frac{1}{\alpha ^{2}}  \label{e21}
\end{equation}
and 
\begin{equation}
3\left( \frac{\dot{\alpha}}{\alpha }\right) ^{2}+3\left( \frac{\dot{\alpha}}{%
\alpha }\right) \left( \frac{\dot{\Phi}}{\Phi }\right) -\frac{\lambda _{1}}{2%
}\left( \frac{\dot{\phi}}{\phi }\right) ^{2}-\frac{\left( 3+2\lambda
_{1}\right) \rho }{\Phi }\equiv -k\,j\frac{B}{\alpha ^{2}}  \label{e22}
\end{equation}
using a notation $j=-1$ for $\omega /2+B<0$ and $j=+1$ for $\omega /2+B>0$.

Thus we find the similar Machian cosmological solution in the generalized
scalar-tensor theory of gravitation with the varying cosmological constant, 
\begin{equation}
\Phi (t)=\zeta \rho (t)t^{2}  \label{e23}
\end{equation}
and 
\begin{equation}
\alpha (t)=bt  \label{e24}
\end{equation}
with 
\begin{equation}
\zeta =1/(\xi -2)\,,  \label{e25}
\end{equation}
\begin{equation}
b=\left\{ 
\begin{array}{l}
(4-\xi ^{2})^{-1/2}\,,\;\;for\;k\,j=-1\;and\;0\leqq \xi <2 \\ 
(\xi ^{2}-4)^{-1/2}\,,\;\;for\;k\,j=+1\;and\;2<\xi \leqq 4\,,
\end{array}
\right.  \label{e26}
\end{equation}
and 
\begin{equation}
B=\frac{\lambda _{1}}{2}-\frac{3}{(\xi -2)(\xi +2)}\,.  \label{e27}
\end{equation}

There is a discontinuity at $\xi =2$ and we restrict to the range $0\leqq
\xi <2$ owing to the evolutionary continuity from $\xi =0$ (the radiation
era) and $\xi =1$ (the mater-dominated era with $p=0$). So we require the
closed model ($k=+1$, $j=-1$) for the attractive gravitational force ($G>0$%
). Both the coupling function $\omega (\phi )=\eta /(\xi -2)$ for $0\leqq
\xi <2$ and the constraint 
\begin{equation}
\frac{\omega (\phi )}{2}+B<0  \label{e31}
\end{equation}
lead necessarily to $\eta >0$ and $\omega <0$. We require $\eta >3$ to avoid
the singularity ($\omega <-3/2$). We get $\omega (\phi )=-\eta /2$ and $%
B=\lambda _{1}/2+3/4$ when $\xi =0$, and so the constraint Eq.(\ref{e31})
gives $\eta -3>2\lambda _{1}$. Experimentally, if we find $\left| \omega
\right| \sim 10^{3}$ \cite{11)}, we obtain $2-\xi \equiv \epsilon \sim
10^{-3}$ (assuming $\eta \approx 3$). Though the constants $\zeta $ and $b$
diverge at $\xi =2$, the scalar field $\phi $ and the expansion parameter $%
a(t)$ themselves do not diverge at $\xi =2$. The crucial point at $\xi =2$
is that the sign of the coupling function $\omega (\phi )$ reverses there ($%
j=-1\rightarrow j=+1$ for $k=+1$).

The expansion parameter is finally expressed as 
\begin{equation}
a(t)\equiv A(\omega )bt=\left[ 6/f(\xi )\right] ^{1/2}t\,,  \label{e28}
\end{equation}
where 
\begin{equation}
f(\xi )\equiv \lambda _{1}(\xi -2)(\xi +2)+\eta (\xi +2)-6\,.  \label{e29}
\end{equation}
When $\xi \rightarrow 2$, we get the finite expansion parameter 
\begin{equation}
a(t)=\sqrt{3/(2\eta -3)}\,t\,.  \label{e30}
\end{equation}
For the enough expansion at present, we need require that the constant $\eta 
$ is not so much. The value $\eta =3$ gives the expansion parameter$\ a(t)=t$
(light velocity). If $\eta >3$ and $\lambda _{1}<0$, the quadratic function $%
f(\xi )$ is always positive for the variable $\xi $ ($0\leqq \xi <2$)\ and
satisfies Eq.(\ref{e31}). The function $f(\xi )$ becomes a maximum at $\xi
_{\max }=-\eta /\lambda _{1}$. In the case $\xi _{\max }\geqq 2$, $f(\xi )$
gives a monotonous increasing function for $0\leqq \xi <2$, and in the case $%
0<\xi _{\max }<2$, gives a monotonous decreasing function for $\xi _{\max
}\leqq \xi <2$. If $\eta >3$ and $\lambda _{1}>0$, the quadratic function $%
f(\xi )$ is positive for $0\leqq \xi <2$ when $\eta -3>2\lambda _{1}$, and
gives always a monotonous increasing function for $0\leqq \xi <2$. If $%
\lambda _{1}=0$, then the function $f(\xi )$ becomes a linear function and
shows a monotonous increase for $\eta >0$. Therefore, if $\lambda _{1}<0$,
and $0\leqq -\eta /\lambda _{1}<2$, the universe shows the slowly
accelerating expansion for the period $-\eta /\lambda _{1}\leqq \xi <2$.

The scalar field $\phi $ for the coupling function $\omega (\phi )=\eta
/(\xi -2)$ is given as the following and we obtain when $\xi \rightarrow 2$%
\begin{equation}
\phi (t)=\frac{8\pi \rho (t)t^{2}}{\left( 3+\lambda _{1}\right) (\xi
-2)+2\eta }\cong \frac{4\pi \rho (t)t^{2}}{\eta }\rightarrow const\,,
\label{e32}
\end{equation}
which converges to a definite and finite constant in the limit. The
asymptotic behavior ($\xi \rightarrow 2$) of the scalar field is the same as
that of the case without the cosmological constant. It should be remarked
that the gravitational constant $G$ is determined by the mass density and
the age of the universe if we adopt $\eta \approx 3$ as the coupling
function is large enough. If $t_{0}=1.5\times 10^{10}\,yr$, we obtain $\rho
_{0}=1.6\times 10^{-29}\,g.cm^{-3}$, which is very near to the critical
density $\rho _{c}\sim 10^{-29}\,g.cm^{-3}$. As the parameter $\xi
\rightarrow 2$, the coupling function $\omega (\phi )$ diverges to the minus
infinity and the gravitational constant approaches dynamically to the
constant $G_{\infty }$

The cosmological constant $\lambda (t)$ decreases rapidly in proportion to $%
t^{-2}$\ as the universe expands and converges to zero when $\xi \rightarrow
2$ ($t\rightarrow +\infty $): 
\begin{equation}
\lambda (t)=\frac{\lambda _{1}}{2}\left( \frac{\dot{\phi}}{\phi }\right)
^{2}\propto \frac{(\xi -2)^{2}}{t^{2}}\rightarrow 0\,.  \label{e33}
\end{equation}
The effective cosmological constant $\Lambda (t)$ introduced in the previous
paper \cite{13)} also decreases rapidly and converges to zero when $\xi
\rightarrow 2$ ($t\rightarrow +\infty $): 
\begin{equation}
\Lambda (t)\equiv -\frac{\omega (\phi )}{2}\left( \frac{\dot{\phi}}{\phi }%
\right) ^{2}\propto \frac{\xi -2}{t^{2}}\rightarrow 0\,.  \label{e34}
\end{equation}
It should be noted that the sign of the cosmological term $\lambda (t)<0$ is
opposite to that of the usual cosmological term in the $(i,i)$ components of
the field equations, and inversely the effective cosmological term $\Lambda
(t)>0$ is opposite in the $(0,0)$ component.

We can estimate the order of each terms appeared in Eqs.(\ref{e12}) and (\ref
{e13}) by means of the present Machian cosmological solution when $\omega
(\phi )=\eta /(\xi -2)\rightarrow -\infty $, for example, 
\begin{equation}
\frac{8\pi }{3+2\left[ \omega (\phi )+\lambda _{1}\right] }\frac{a^{2}\xi
\rho }{\phi }\sim \xi -2\rightarrow 0\,,  \label{e35}
\end{equation}
\begin{equation}
\frac{8\pi a^{2}\rho }{\phi }\rightarrow \frac{6\eta }{2\eta -3}\sim O(1)\,,
\label{e36}
\end{equation}
\begin{equation}
\frac{8\pi a^{2}p}{\phi }\rightarrow -\frac{2\eta }{2\eta -3}\sim O(1)\,,
\label{e37}
\end{equation}
\begin{equation}
\frac{\omega (\phi )}{2}a^{2}\left( \frac{\dot{\phi}}{\phi }\right) ^{2}\sim
\xi -2\rightarrow 0\,,  \label{e38}
\end{equation}
\begin{equation}
\frac{\lambda _{1}}{2}a^{2}\left( \frac{\dot{\phi}}{\phi }\right) ^{2}\sim
(\xi -2)^{2}\rightarrow 0\,,  \label{e39}
\end{equation}
\begin{equation}
a\dot{a}\left( \frac{\dot{\phi}}{\phi }\right) \sim \xi -2\rightarrow 0\,,
\label{e40}
\end{equation}
and 
\begin{equation}
a^{2}\frac{\ddot{\phi}}{\phi }\sim \xi -2\rightarrow 0\,.  \label{e41}
\end{equation}
The crucial difference from the correspondences \cite{14)} in the
Brans-Dicke theory \cite{15)} is that even the term Eq.(\ref{e38}) converges
to zero when $\omega (\phi )\rightarrow -\infty $. This means that the
abnormal term vanishes and the generalized scalar-tensor theory of
gravitation reproduces the correspondent solution of general relativity with
the same energy-momentum tensor completely when $\xi \rightarrow 2$ ($%
t\rightarrow +\infty $).

In the Brans-Dicke theory, the coupling parameter $\omega $ is arbitrary,
and so we realize that the Machian solution does not reduce to that of
general relativity for the fixed finite time $t$ when $\left| \omega \right|
\rightarrow +\infty $. However, we cannot determine, for example, the limit
of the term $\Lambda (t)\propto \omega /t^{2}$ when $\left| \omega \right|
\rightarrow +\infty $ and $t\rightarrow +\infty $ in the Brans-Dicke theory.
If we regard the coupling function $\omega (\phi )$ as an arbitrary
parameter in the generalized scalar-tensor theory, we obtain the similar
correspondences (for example, $\omega (\phi )a^{2}\left( \dot{\phi}/\phi
\right) ^{2}\sim O(1)$) in the Brans-Dicke theory for the same Machian
cosmological solution for the fixed finite time $t$. In this case, the
generalized scalar-tensor theory does not reduce to general relativity when $%
\left| \omega \right| \rightarrow +\infty $ and $\omega ^{-3}d\omega /d\phi
\rightarrow 0$. We find that the scalar field $\phi $ converges to zero when 
$\left| \omega \right| \rightarrow +\infty $ according to the postulate $%
\phi =O(\rho /\omega )$. For the present Machian cosmological solution in
the generalized scalar-tensor theory, we know the behavior of the coupling
function $\omega (\phi )$ when $t\rightarrow +\infty $ and so we can
estimate definitely the limit of the solution when $t\rightarrow +\infty $.
Thus the generalized scalar-tensor theory of gravitation approaches
dynamically to general relativity in the result of the evolution of the
universe when $t\rightarrow +\infty $.

Let us remember the Friedmann equations with the cosmological term for the
homogeneous and isotropic universe in general relativity: 
\begin{equation}
2a\ddot{a}+\dot{a}^{2}+k-\lambda a^{2}=-\kappa pa^{2}\,,  \label{e42}
\end{equation}
\begin{equation}
\frac{3}{a^{2}}\left( \dot{a}^{2}+k\right) -\lambda =\kappa \rho \,,
\label{e43}
\end{equation}
where $\kappa $ is Einstein's gravitational constant. We get from Eq.(\ref
{e42}) when $\ddot{a}=0$ and $p=-\rho /3$%
\begin{equation}
\frac{3}{a^{2}}\left( \dot{a}^{2}+k\right) -3\lambda =\kappa \rho \,,
\end{equation}
and find $\lambda =0$. If we require $\ddot{a}(t)=0$ and $\lambda
(t)\rightarrow 0$ ($t\rightarrow +\infty $), we obtain $p\rightarrow -\rho
/3 $. This fact supports that the final state of the Machian cosmological
solution with the decaying cosmological constant is the negative pressure $%
\gamma =-1/3$ ($\xi =2$). Taking Eqs.(\ref{e30}) and (\ref{e35})-(\ref{e41})
($k=+1$), and $\kappa \equiv 8\pi G_{\infty }$ into account, we observe that
Eqs.(\ref{e12}) and (\ref{e13}) reduce to Eqs.(\ref{e42}) and (\ref{e43})
respectively when $\omega (\phi )\rightarrow -\infty $ ($t\rightarrow
+\infty $). The constant $\eta $ still remains indefinite. It is interesting
that the Machian cosmological solution in the Brans-Dicke theory, of which
the coupling parameter is constant and arbitrary, is correspondent to the
static Einstein universe ($\dot{a}\rightarrow 0$ when $\omega \rightarrow
-\infty $).

The cosmological constant $\lambda (\phi )$ is not consistent with the
Machian cosmological solution satisfying $\phi =O(\rho /\omega )$ for the
variable parameter $\xi $. Only the present form of the cosmological
constant $(1/2)\lambda _{1}\phi _{,\,\mu }\phi ^{,\,\mu }/\phi ^{2}$ admits
the Machian solution in the generalized scalar-tensor theory of gravitation.
The constraint $\omega (\phi )<-3/2$ leads to $\eta >3$. If $\lambda _{1}<0$
and $0\leqq -\eta /\lambda _{1}<2$, the universe surely shows the slowly
accelerating expansion for the later period from $\xi =-\eta /\lambda _{1}$,
though the expansion parameter is explicitly a linear function of $t$. For
example, if we adopt $\eta =3$ and $\lambda _{1}=-3$, the universe exhibits
the slowly decelerating expansion before the matter-dominated era $\xi =1$
(for the positive pressure), the slowly accelerating expansion after the
matter-dominated era (for the negative pressure) respectively.

In the generalized scalar-tensor theory of gravitation, not only the
gravitational constant but also the variation of the coupling function is
derived from the evolution of the universe. The remaining problem is to
determine the time-variation of the parameter $\xi (t)$ by the physical
evolution of matter in the universe. Our conjecture is that the universe
passed the radiation era ($\xi =0$) and the matter-dominated era with
negligible pressure ($\xi =1$) rapidly in the early stage, and has been
staying the negative pressure era ($1<\xi <2$, maybe almost near $\xi =2$)
for the almost all period of $10^{10}yr$. Finally, the universe will
approach to the state of $\xi =2$ ($\gamma =-1/3$) as it expands for ever.%
\newline
\newline
\textbf{Acknowledgment}

The author is grateful to Professor Carl Brans for helpful discussions and
his hospitality at Loyola University (New Orleans) where this work was done.
He would also like to thank the Nagasaki Prefectural Government for
financial support.


\begin{thebibliography}{99}
\bibitem{1)}  P.G.Bergmann, Int. J. Theor. Phys. 1, 25 (1968).

\bibitem{2)}  R.V.Wagoner, Phys. Rev. D1, 3209 (1970).

\bibitem{3)}  K.Nordtvedt, Astrophys. J. 161, 1059 (1970).

\bibitem{4)}  A.Burd and A.Coley, Phys. Lett. B267, 330 (1991).

\bibitem{5)}  J.D.Barrow and J.P.Mimoso, Phys. Rev. D50, 3746 (1994).

\bibitem{6)}  J.P.Mimoso and D.Wands, Phys. Rev. D52, 5612 (1995).

\bibitem{7)}  A.Serna and J.M.Alimi, Phys. Rev. D53, 3074 (1996).

\bibitem{8)}  J.D.Barrow and P.Parsons, Phys. Rev. D55, 1906 (1997).

\bibitem{9)}  A.Billyard, A.Coley, and J.Ibanez. Phys. Rev. D59, 023507
(1999).

\bibitem{10)}  A.Miyazaki, gr-qc/0102105, 2001.

\bibitem{11)}  X.Chen, M.Kamionkowski, Phys. Rev. D60, 104036 (1999).

\bibitem{12)}  S.Perlmutter, M.S.Turner, and M.White, Phys. Rev. Lett. 83,
670 (1999).

\bibitem{13)}  A.Miyazaki, Nuovo Cimento 68B, 126 (1982).

\bibitem{14)}  A.Miyazaki, gr-qc/0012104, 2000.

\bibitem{15)}  C.Brans and R.H.Dicke, Phys. Rev. 124, 925 (1961).
\end{thebibliography}
\end{document}